\documentclass[traditabstract]{aa}

\usepackage{amsmath, amssymb, upgreek}
\usepackage{graphicx}
\usepackage{txfonts}
\usepackage{natbib}

\def\vv{\varv}
\def\erf{{\mathrm{erf}}}

\begin{document}

\title{Surface flux evolution constraints for flux transport dynamos}

\author{R.~H.~Cameron\inst{1} \and D.~Schmitt\inst{1} \and J.~Jiang\inst{2} \and E.~I\c{s}{\i}k\inst{3}}

\institute{Max-Planck-Institut f\"ur Sonnensystemforschung, Max-Planck-Str. 2,
37191 Katlenburg-Lindau, Germany, \email{cameron@mps.mpg.de}
\and Key Laboratory of Solar Activity, National Astronomical Observatories,
Chinese Academy of Sciences, Beijing 100012, China
\and Department of Physics, Faculty of Science \& Letters, Istanbul K\"ult\"ur
University, Atak\"oy Campus, Bak\i rk\"oy 34156, Istanbul, Turkey}

\date{Received ; accepted}

\abstract{The surface flux transport (SFT) model of solar magnetic fields
involves empirically well-constrained velocity and magnetic fields. The basic
evolution of the Sun's large-scale surface magnetic field is well described by
this model. The azimuthally averaged evolution of the SFT model can be compared
to the surface evolution of the flux transport dynamo (FTD), and the evolution
of the SFT model can be used to constrain several near-surface properties of
the FTD model.

We compared the results of the FTD model with different upper boundary
conditions and diffusivity profiles against the results of the SFT model. Among
the ingredients of the FTD model, downward pumping of magnetic flux, related to
a positive diffusivity gradient, has a significant effect in slowing down the
diffusive radial transport of magnetic flux through the solar surface. Provided
the pumping was strong enough to give rise to a downflow of a magnetic Reynolds
number of 5 in the near-surface boundary layer, the  FTD using a vertical
boundary condition matches the SFT model based on the average velocities above
the boundary layer. The FTD model with a potential field were unable to match
the SFT results.}

\authorrunning{Cameron et al.}
\titlerunning{}
\maketitle

\section{Introduction}

The flux transport dynamo model (FTD) attempts to explain the large-scale
evolution of the Sun's magnetic field. The central ideas behind the model are
that poloidal flux is wound up by differential rotation until it becomes
sufficiently strong that magnetic buoyant flux tubes emerge through the solar
surface. The erupted field is in the form of a bipolar active region, and the
two opposite polarities are observed to be systematically tilted with respect
to the equator (Joy's law). This tilt is such that the leading polarity is
slightly closer to the equator than the following polarity. This latitudinal
offset means that poloidal field has been created from the toroidal flux and,
in the language of dynamo theory, the emergence process is a non-local alpha
effect \citep{Kitchatinov11b}. The poloidal flux is then stretched and diffused
by surface motions, reversing the polar fields and completing one half of a
solar cycle. For a review of the basic ideas, see \citet{Charbonneau10}. This
picture has recently gained observational support from the analysis by
\citet{Dasi-Espuig10} and later by \citet{Kitchatinov11a}, which show that the
observed sunspot group tilt angles, which go into the construction of the
poloidal source term, vary systematically from cycle to cycle in a way which
possibly can explain the observed changes in cycle amplitudes during the
twentieth century.

The winding up of the field by differential rotation and the rise of the tubes
to the surface are hidden below the photosphere. The evolution of the field
after it has broken through the surface can be and has been observed. The
surface flux transport model (SFT) has been found to provide a good description
of the large-scale evolution after emergence. For a detailed historical
account, see \citet{Sheeley05}. This model assumes that the magnetic field is
purely radial at the surface and evolves passively driven by surface flows
including differential rotation, meridional circulation and small-scale
convective motions (granulation and supergranulation). The small-scale motions
essentially cause the magnetic field to undergo a random walk and hence can be
treated as a diffusive term. The ingredients which go into the SFT model are
all observable, as is the output of the model -- it is thus tightly constrained
and supported by observations. For example, \citet{Cameron10} showed that the
SFT model, with the observed cycle-to-cycle variations of the tilt angle, can
reproduce the inferred open magnetic flux of the Sun. Since the open flux
during the maxima and minima of activity reflect the equatorial and axial
dipole moments respectively, the model's ability to reproduce the open flux
over an extended period is a strong test of the model.

In this paper we investigate what constraints can be inferred for the FTD
model, given that it should also reproduce the same surface dynamics as is
described by the SFT model. We have used the FTD code developed at the 
Max-Planck-Institut f\"ur Sonnensystemforschung. For
the SFT model we have used a 1-D surface flux transport model developed at the
MPS. The 1-D SFT model includes exactly the component which can be compared
between the two models. The details of the two approaches will be discussed in
Sect.~2. In Sect.~3 we present the results of the simulations and compare the
two models. The effect of varying some of the most important unconstrained
parameters and the boundary condition will be discussed in Sect.~4. We will
conclude in Sect.~5 with the finding that the appropriate boundary condition
for FTD models is that the field is vertical at the surface, and that a certain
amount of turbulent pumping must be included for the FTD simulations to mimic
the surface behavior of the SFT model and to thus match the observations.

Downward pumping has in particular been discussed for the base of the solar
convection zone. Direct numerical simulations show a downward transport of
large-scale magnetic field near the base of convective unstable layers
\citep[e.g.,][]{Jennings92,Tobias98,Tobias01,Ossendrijver02} though it is not clear
whether this should be interpreted in terms of turbulent pumping
\citep{Zeldovich57,Raedler68} or of topological pumping \citep{Drobyshevski74}.
In mean-field dynamo models of the solar cycle turbulent pumping is often not
included. In those cases where it is included, most of the attention is focussed
on its role in transporting flux into the top of the stable layer immediately
below the convection zone 
\citep[see for example][]{Brandenburg92, Kaepylae06, DoCao11,Kitchatinov11b,Kitchatinov12}. 
The effect of pumping throughout the convection zone was considered by 
\citep{Guerrero08}, who found that the pumping affects 
whether the preferred mode of the solution is dipolar or quadrupolar, and identified
the possible importance of radial transport by the pumping in the dynamo process.   
In this paper we are especially
paying attention to the pumping in the near-surface boundary layer
\citep{Miesch11} which is necessary to match FTD to SFT simulations of the
magnetic flux on the solar surface.

\section{Physical models and numerical codes}

\subsection{The Flux Transport Dynamo (FTD) model}

The flux transport dynamo equations describe the induction, advection and
diffusion of a large-scale magnetic field. Their axisymmetric form is:
\begin{eqnarray}
\frac{\partial A}{\partial t} &=& \eta(r)\left(\nabla^2-\frac{1}{(r\sin \theta)^2}\right)A \nonumber\\
 & &-\frac{{\vec{u}}_{\mathrm{m}}(r,\theta) + {\vec{u}}_{\mathrm{p}}(r,\theta)}{r\sin\theta}\cdot\nabla \left(A r \sin \theta \right)
   +\alpha (B)
\label{eqn:A} \\
\frac{\partial B}{\partial t} &=& \eta(r)\left(\nabla^2-\frac{1}{(r\sin \theta)^2}\right)B
   + \frac{1}{r}\frac{\partial \eta}{\partial r} \frac{\partial rB}{\partial r} \nonumber \\
   & & {}- r \sin\theta \left({\vec{u}}_{\mathrm{m}}(r,\theta) +{\vec{u}}_{\mathrm{p}}(r,\theta) \right)\cdot\nabla
   \left(\frac{B}{r \sin \theta}\right) \nonumber\\
   & &- B\,\nabla \cdot \left(\vec{u}_{\mathrm{m}}(r,\theta) + \vec{u}_{\mathrm{p}}(r,\theta)\right) \nonumber \\
   & & {}+ r \sin \theta \left(\nabla \times \left(A {\vec{\hat e}}_\phi\right)\right) \cdot \nabla \Omega(r,\theta)
\label{eqn:B}
\end{eqnarray}
where $A(r,\theta)$ is the $\phi$-component of the vector potential associated
with the poloidal components of ${\vec{B}}$,  $B(r,\theta)$ is the toroidal
component of the field, ${\vec{u}}_{\mathrm{m}}(r,\theta)$ is the velocity in the meridional plane, 
$\Omega(r,\theta)$ is the angular velocity,
$\vec{u}_{\mathrm{p}}(r,\theta)$ is a velocity field corresponding
to the pumping of the magnetic field
and $\alpha$ is a source term in the equation
for $A$ corresponding to the generation of poloidal flux from toroidal flux.
Since the purpose of the current study is to compare the response of the 
SFT and FTD models to equivalent sources of poloidal flux, we restrict ourselves
to the case $\alpha=0$ -- for other choices of $\alpha$ we would need to modify the
source term in the SFT model accordingly.
In relation to the term $\vec{u}_{\mathrm{p}}$ it is important to note that, 
as in \cite{Guerrero08}, it does not correspond 
to a true motion of the fluid and need not satisfy $\nabla \cdot \rho \vec{u}_{\mathrm{p}}=0$.
Rather it is a parametrization of the effect of the turbulent motions on the field: 
for diamagnetic pumping it has the form $\vec{u}_{\mathrm{p}}(r,\theta)=-\frac{1}{2}\nabla \eta$. 
Other effects, such as topological pumping, are also expected to transport the field downwards,
and for this study we assume that the combined effects of the turbulent convection,
including diamagnetic pumping, can be written in the form 
$\vec{u}_{\mathrm{p}}(r,\theta)=-\frac{k}{2}\nabla \eta$, with $k \ge 1$. This choice allows us to 
vary the magnitude of the pumping in the near surface layers.

We solve the dynamo equations (\ref{eqn:A}) and (\ref{eqn:B}) forward in time
in a spherical shell $r_0\le r\le R_{\sun}$ with inner boundary
$r_0=0.65R_{\sun}$ matching to a perfect conductor and outer boundary matching
to either a radial field or vacuum conditions outside. This leads to the boundary conditions
\begin{eqnarray}
A=0 \quad \mathrm{and} \quad \frac{\partial}{\partial r}(rB)=0 \quad \mathrm{at} \quad r=r_0
\end{eqnarray}
and
\begin{eqnarray}
\frac{\partial}{\partial r}(rA)=0 \quad \mathrm{and} \quad B=0 \quad \mathrm{at} \quad r=R_{\sun}
\end{eqnarray}
for the field to be vertical at the Sun's surface, or alternatively
\begin{eqnarray}
A&=&\sum_k a_kP_k^1(\cos\theta),\\
 \frac{\partial A}{\partial r}&=&-\sum_k (k+1)a_kP_k^1(\cos\theta)
   \quad \mathrm{and} \\
 B&=&0 \quad \mathrm{at} \quad r=R_{\sun}
\end{eqnarray}
for matching to a potential field outside. At the poles we require regularity
resulting in
\begin{eqnarray}
A=B=0 \quad \mathrm{at} \quad \theta=0,\pi \;.
\end{eqnarray}
The equations are discretized using second order accurate centered
finite differences on an equidistant grid and forwarded in time
with an Alternating Direction Implicit scheme for the diffusion terms and an
explicit scheme for the induction and advection terms. The code is tested
against the dynamo benchmark of \citet{Jouvre08}.

For current purposes we will consider $\alpha=0$ so that there is no source of
poloidal field during the simulation. From any initial condition the field must
then eventually decay towards zero, however at any finite time the magnetic
field will depend on the initial field and can be compared with the result of
the SFT model.

For the initial condition we take
\begin{eqnarray}
A &=& \frac{1}{8}R_{\sun}\left(1+\erf\left(\frac{r-r_1}{\Delta r}\right)\right) \times
  \left(1+\erf\left(\frac{\theta-\theta_1}{\Delta \theta}\right)\right) \times \nonumber \\
  & & \hspace{1cm}  \left(1-\erf\left(\frac{\theta-\theta_2}{\Delta \theta}\right)\right), \\
  B&=&0,
\end{eqnarray}
where $r_1=0.80 R_{\sun}$, $\theta_1=80^{\circ}$, $\theta_2=86^{\circ}$,
$\Delta \theta=2.9^{\circ}$ and $\Delta r=0.01 R_{\sun}$. This corresponds to
an isolated bipole emerging on the solar surface slightly north of the equator.
Since both the SFT and FTD studied here are linear, the evolution of such a
bipole is independent of the emergence and evolution of other emerging groups.
To check that the models are consistent, it is thus sufficient to follow the
evolution of a single feature starting near the equator to the poles.

The velocity in the meridional plane is taken from \citet{Dikpati04}. The
velocity components can be written in terms of a stream function as
\begin{eqnarray}
{\vec{u}}_{\mathrm{m}}(r,\theta)&=& \frac{\vv_0}{\rho} \frac{1}{r\sin\theta} \frac{\partial\Psi\sin\theta}{\partial\theta}\,{\vec{\hat e}}_r
   -\frac{\vv_0}{r \rho} \frac{\partial r \Psi}{\partial r}\,{\vec{\hat e}}_\theta \;,
\label{eqn:ur_start}
\end{eqnarray}
where
\begin{eqnarray}
\xi &=& \frac{R_{\sun}}{r}-0.985 \;, \\
\rho &=&\xi^m
\end{eqnarray}
and
\begin{eqnarray}
\Psi(r,&\theta&) = \frac{R_{\sun}}{r}  \times  \nonumber\\
   & & \left( \frac{-1}{m+1}\xi^{{m+1}}
   +\frac{c_1}{2m+1}\xi^{2m+1}-\frac{c_2}{2m+p+1}\xi^{2m+p+1} \right)  \nonumber\\
   & &\times \sin^{q+1}\theta\cos\theta
\end{eqnarray}
with
\begin{eqnarray}
c_1 = \frac{(2m+1)(m+p)}{(m+1)p} \xi_0^{-m} \;,
\end{eqnarray}
\begin{eqnarray}
c_2 = \frac{(2m+p+1)m}{(m+1)p}\xi_0^{-(m+p)} \;.
\label{eqn:ur_end}
\end{eqnarray}
For the reference case we take $q=1.5$, $m=1.5$, $p=3$, $r_0=0.7$ and
$\xi_0=\xi(r_0)$. In all cases $\vv_0$ is chosen so that the maximum meridional
velocity at $r=R_{\sun}$ is 15 m/s. The resulting velocity approximates the
meridional circulation in the solar convection zone derived by numerical
modeling of \citet{Rempel05} and \citet{Kitchatinov05} and is consistent with
the velocity in the subsurface layers as derived from helioseismology 
\citep[as measured, e.g. by][]{Giles97}.

The differential rotation is taken from \citet{Belvedere00}, and is also used
e.g. by \citet{Kitchatinov11a},
\begin{eqnarray}
\Omega(r,\theta) = \sum_{j=0}^2 \cos\left(2j\left(\frac{\pi}{2}-\theta\right)\right)\,\sum_{i=0}^4 c_{ij} r^i
\end{eqnarray}
where the coefficients $c_{ij}$ are given in Table 1 of \citet{Belvedere00}.
This approximates the internal rotation of the Sun as derived from
helioseismological inversions \citep[as reported, e.g., by][]{Schou98}.


For the diffusivity we assumed
\begin{eqnarray}
\eta(r) &=&\eta_0 +\frac{\eta_1-\eta_0}{2}\left(1+\erf\left(\frac{r-0.7R_{\sun}}{0.02 R_{\sun}}\right)\right)  \nonumber\\
    & &+\frac{\eta_2-\eta_1}{2}\left(1+\erf\left(\frac{r-0.95R_{\sun}}{0.02 R_{\sun}}\right)\right)
 \label{eqn:eta}
\end{eqnarray}
with $\eta_0=0.1$~km$^2$s$^{-1}$, $\eta_1=10$~km$^2$s$^{-1}$ and
$\eta_2=250$~km$^2$s$^{-1}$, see e.g. \citet{Munoz-Jaramillo11}. Here $\eta_2$
represents the turbulent diffusivity in the near-surface layers, $\eta_1$ in
the bulk of the convection zone, and $\eta_0$ in the overshoot region at the
base of the convection zone. Other choices will be considered in Sect.~4.

\begin{figure}[h]
  \includegraphics[scale=0.5]{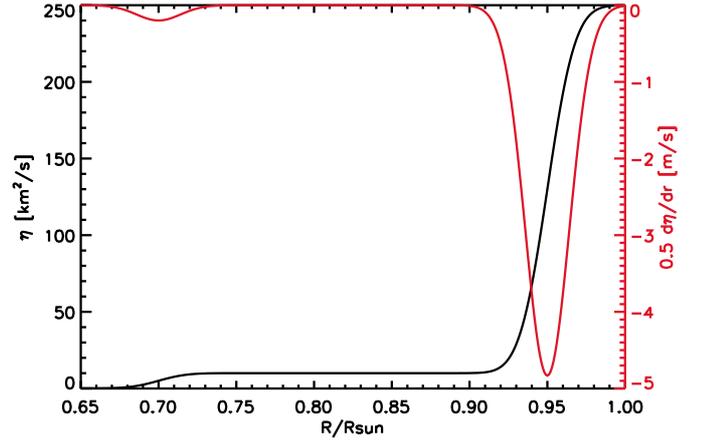}
  \caption{The assumed profile of the turbulent diffusivity is shown in black, the effective radial velocity
      due to the radial derivative of the turbulent diffusivity, for the case $k=1$, is shown in red.}
  \label{fig:eta}
\end{figure}

Recently \citet{Kitchatinov11b} have highlighted the importance of downward
pumping of magnetic fields due to gradients in the turbulent diffusivity, and
have argued that this is particularly important near the base of the convection
zone. We here consider downward pumping in the near-surface layers. We have
found it necessary to increase the strength of the downward pumping from its
usual value of $(1/2) (\partial\eta/\partial r)$ in order to obtain a match
between the FTD and SFT models. We have therefore introduced in
Eqs.~(\ref{eqn:A}) and (\ref{eqn:B}) a scaling factor $k$ which we have varied
between 0 and 20. The diffusivity profile and the corresponding diamagnetic 
pumping velocity with $k=1$ is shown in Fig.~\ref{fig:eta}.

While we solve both Eqs.~(\ref{eqn:A}) and (\ref{eqn:B})
we note that the comparison with the SFT model only depends on 
Eq.~(\ref{eqn:A}) as $\alpha=0$.
The physical ingredients which affect $A$ are the meridional flow, the
radial and latitudinal diffusion, and the downward pumping.  


\subsection{The Surface Flux Transport  (SFT) model}

The SFT model, which describes the evolution of the magnetic field on the solar
surface, assumes that the field is vertical and evolves passively under the
action of the surface flows. The surface differential rotation and surface
meridional flow towards the pole are modeled as systematic flows, while
granular and supergranular flows are assumed to only cause the fields to
diffuse across the solar surface. In this sense correlations between the radial
component of the magnetic field $B_r$ and the supergranular velocity field
$U_{\mathrm{SG}}$ are ignored, i.e. it is assumed that $\langle U_{\mathrm{SG}} B_r
\rangle=\langle U_{\mathrm{SG}}\rangle \langle B_r \rangle$, and since
differential rotation and the meridional flow have been removed, $\langle
U_{\mathrm{SG}}\rangle=0$. This assumption is not justified since the magnetic
field and supergranular velocity fields are correlated, as the magnetic field
is located at the edge of the supergranules. This presumably accounts for the
observation by \citet{Meunier05} that magnetic fields rotate faster than the
local plasma, with the extent of the prograde motion depending on the technique
used to measure the velocity. In the current context this is a small effect
which can be ignored.

The SFT model additionally assumes that there is no transport of flux, either
advective or diffusive, across the solar surface. The relevant equation is
\begin{eqnarray}
\frac{\partial B_r}{\partial t} &=&
   - \omega(\theta)\frac{\partial B_r}{\partial\phi}
   - \frac{1}{R_{\sun}\sin\theta}\frac{\partial}{\partial\theta}\left[\vv(\theta)B_r \sin\theta \right] \nonumber \\
   & & {}+ \frac{\eta}{R_{\sun}^2}\left[\frac{1}{\sin \theta}\frac{\partial}{\partial\theta}
   \left(\sin\theta\frac{\partial B_r}{\partial\theta}\right)
   + \frac{1}{\sin^2\theta}\frac{\partial^2 B_r}{\partial\phi^2}\right]
\end{eqnarray}
where $B_r$ is the radial component of the magnetic field, $\theta$ is the
heliographic colatitude, and $\phi$ is the heliographic longitude.
$\omega(\theta)$ is the surface differential rotation and $\vv(\theta)$ is the
surface meridional flow. For the purposes of comparison with the  FTD
simulation, we take $\vv(\theta)={\vec{u}}_{\mathrm{m}}(R_{\sun},\theta)\cdot{\vec{\hat
e}}_\theta$, $\omega(\theta)=\Omega(R_{\sun},\theta)$, and
$\eta=\eta(R_{\sun})=250$~km$^2$/s.

For comparison with the FTD simulation, we can only use the azimuthally
averaged (signed) field strength. This averaged field is independent of the
initial structure of the field in the azimuthal direction and hence we can take
\begin{eqnarray*}
B_r(\theta)=\frac{1}{r\sin\theta}\frac{\partial}{\partial\theta}(A\sin\theta)
\end{eqnarray*}
as our initial condition, consistent with the initial condition of the FTD
simulation. The solution to this one dimensional problem, $B_r(\theta,t)$, can
be directly compared to $R_{\sun}(\partial/\partial\theta)(A\sin\theta)$ from
the FTD simulation. We have used the code described in \citet{Cameron07} to
solve this 1-D surface flux transport problem.

\section{Reference case}

\begin{figure}[h]
  \includegraphics[scale=0.7]{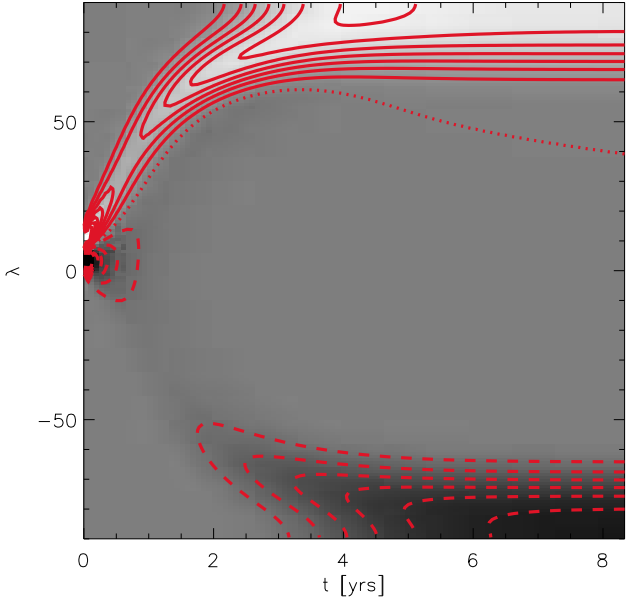}
  \caption{Evolution of the azimuthally averaged signed field strength from the
    SFT simulation, with black and white representing opposite polarities saturated
    at 36\% of the initial azimuthally averaged field strength. The solid and dashed 
   red contours indicate where the field strength reaches $\pm 5$\%, $\pm 10$\%, etc 
   of its maximum value, with the dotted curve representing the 0 level.}
  \label{fig:ft}
\end{figure}

\begin{figure}[h]
  \includegraphics[scale=0.5]{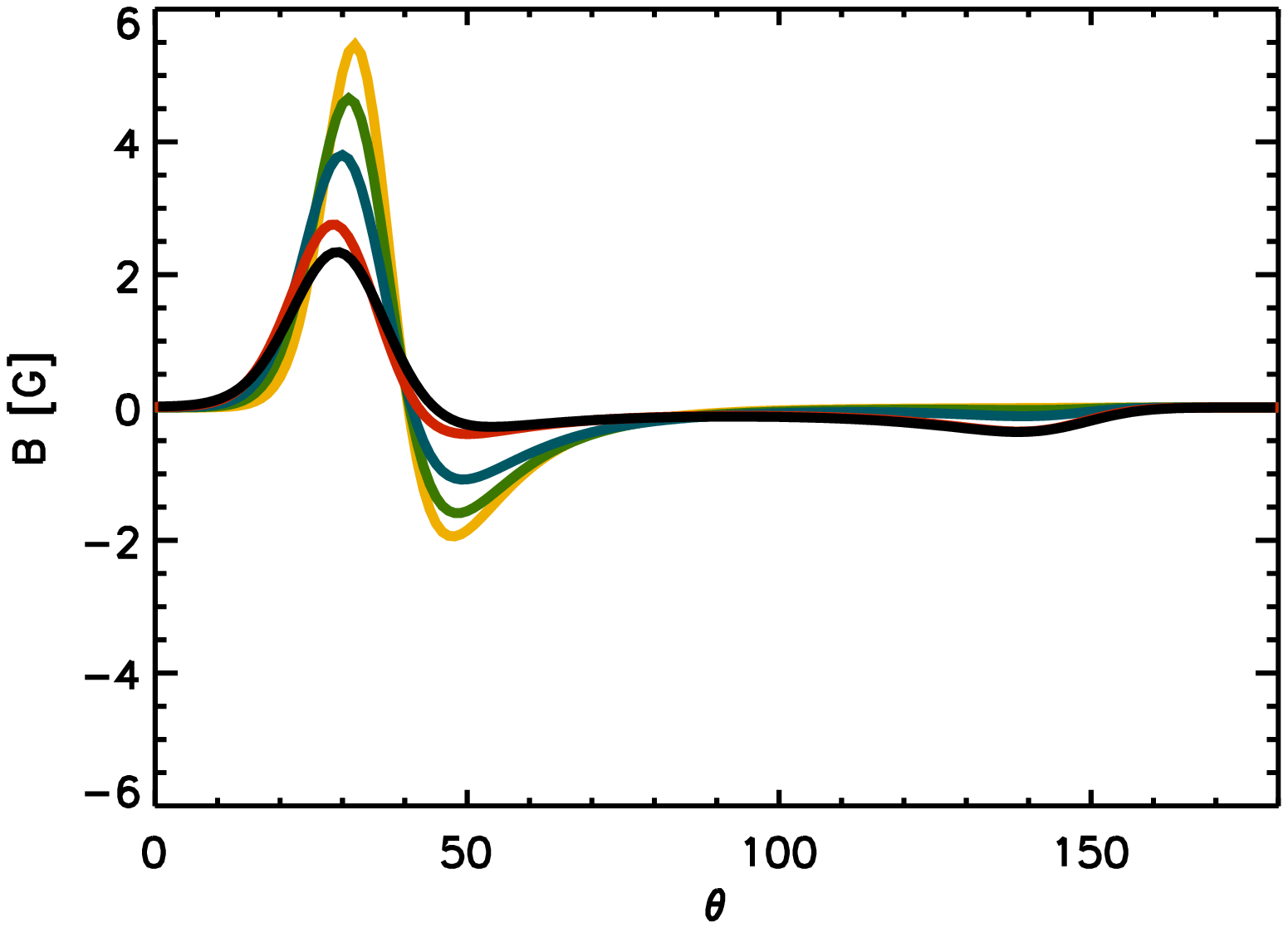}

  \includegraphics[scale=0.5]{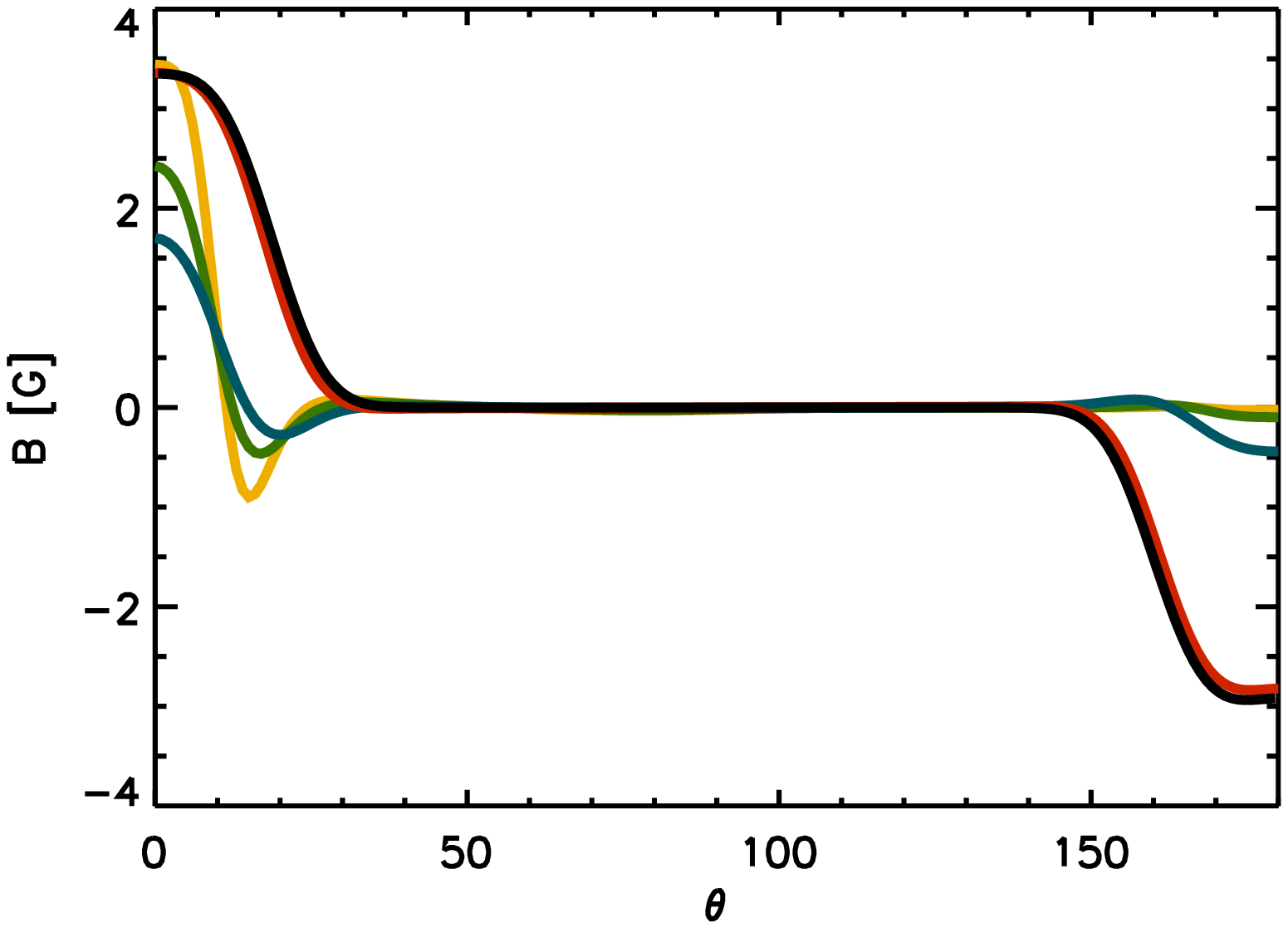}
  \caption{Azimuthally averaged signed field from the SFT model (black) vs
   surface field from FTD simulations at $t=18$ months (top)
   and $t=72$ months (bottom) with a vertical
   field outer boundary and pumping with factors $k=5$ (red),
   2 (blue), 1 (green), and 0 (yellow).}
  \label{fig:vert}
\end{figure}

In Fig.~\ref{fig:ft} the evolution of the surface flux according to the SFT
model is displayed. Figure~\ref{fig:vert} shows the surface latitudinal
dependence of the different FTD models with the vertical boundary condition
with that from the SFT model shown for comparison. We note that for
$k\gtrsim5$, both FTD and SFT models match very well. For $k=0$ the match is
much worse, e.g. there is too little flux in the southern hemisphere ($\theta > 90$)
at $t=72$ months. In the northern hemisphere at $t=18$ months, the amplitude of
the field in the FTD model is greater for both polarities. By $t=72$ months the
amplitude of the field in the southern hemisphere has also fallen as the
opposite polarities are merging. This implies that downward pumping
corresponding to at least $k=5$ is required for the FTD model to be consistent
with the SFT model and therefore with observations.

\begin{figure}[h]
  \includegraphics[scale=0.9,angle=90]{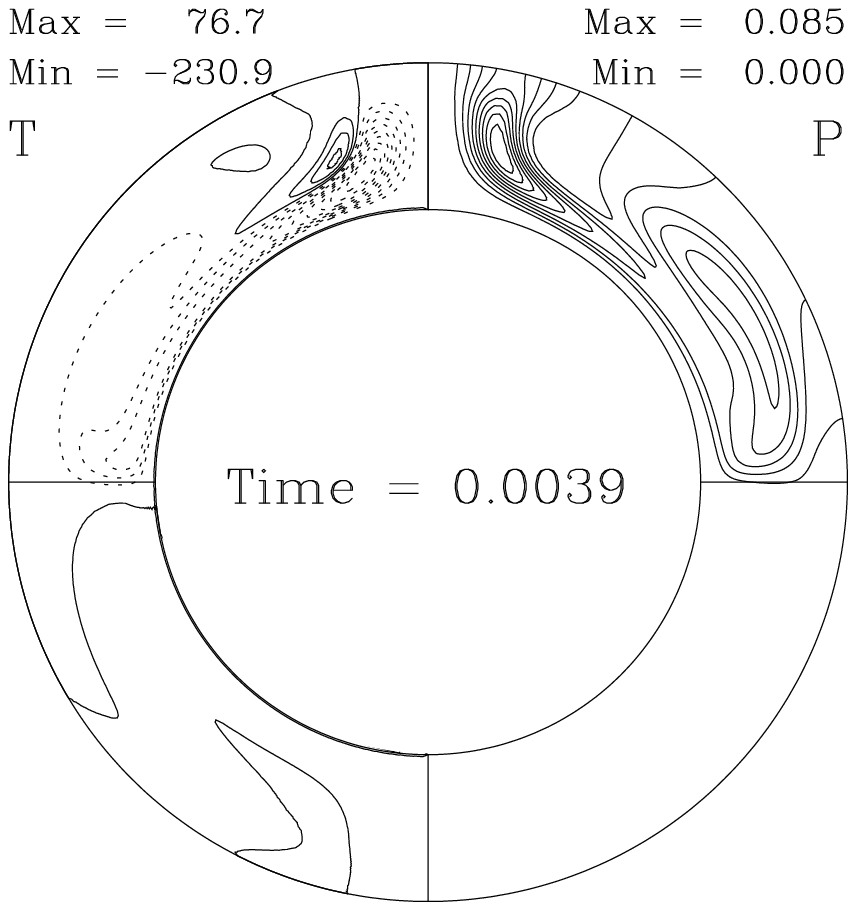}

  \includegraphics[scale=0.9,angle=90]{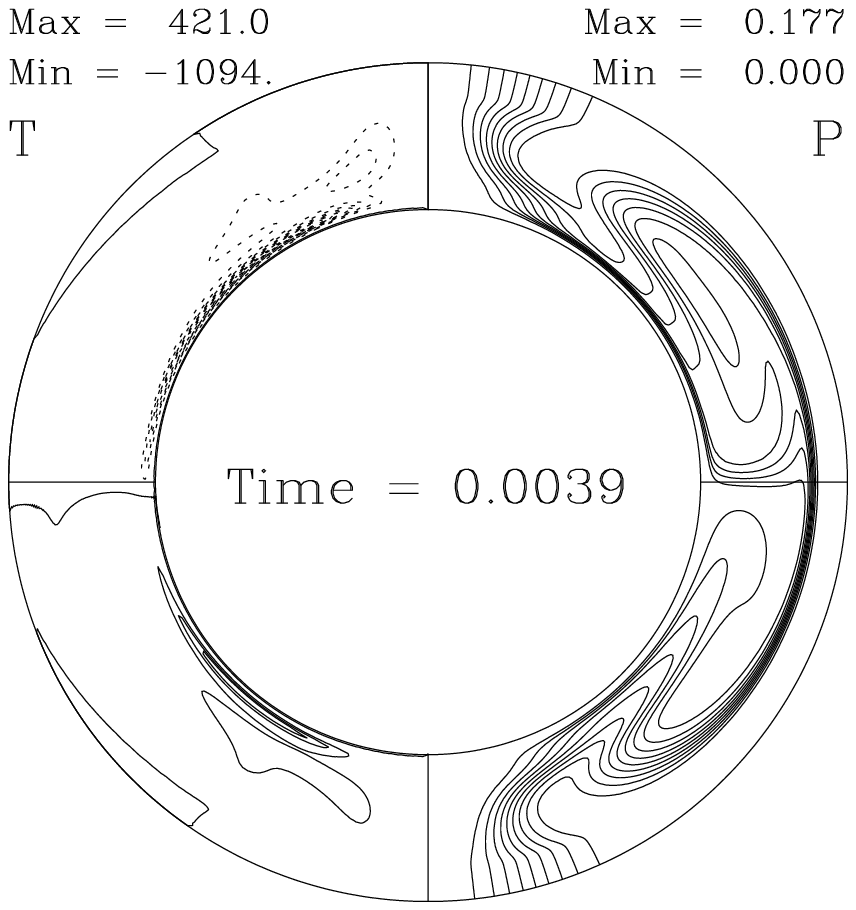}
  \caption{Magnetic field structure from the FTD simulations at $t=72$ months for the case with
          a vertical boundary condition and $k=0$ (top) and $k=5$ (bottom). In each subpanel
          the left half shows contours of the toroidal field ($T$), the right panel shows selected fieldlines of the poloidal field
          (formally it shows contours of $P=r \sin \theta A(r,\theta)$). The dashed contours of the toroial field indicate
           negative fields, the solid contours represent either zero or positive toroidal field. In particular
           the solid contours which touch the boundaries correspond to zero toroidal flux.}
  \label{fig:fl_vert}
\end{figure}

The reason why downward pumping is important can be seen in
Fig.~\ref{fig:fl_vert} where, in the case without pumping, the diffusive
emergence of flux through the upper boundary is obvious. This emergence of flux
is strongly inhibited by the downward pumping. The requirement that the
downward pumping should inhibit the diffusion of flux across the surface is
captured by the corresponding magnetic Reynolds number $R_m$ being larger than
1:
\begin{eqnarray*}
R_m&=&\frac{|{\vec{u}}_\mathrm{p}|L}{\eta} \nonumber\\
   &=&\frac{(k/2)(\partial\eta/\partial r)L}{\eta} \nonumber\\
   &\approx&\frac{(k/2)[(\eta_2-\eta_1)/L]L}{(\eta_2+\eta_1)/2}\nonumber\\
   &\approx& k\nonumber\\
\quad \mathrm{for} \quad \eta_1\ll\eta_2 \;.
\end{eqnarray*}
Here ${\vec{u_{\mathrm{p}}}}$ is the pumping velocity and $L$ is the boundary layer
thickness corresponding to the region over which $\eta$ changes from its value
throughout the bulk of the convection zone $\eta_1$ to its surface values
$\eta_{2}$. Basing $R_m$ on the mean over this transition yields $R_m\approx k$
when $\eta_1\ll\eta_2$. To prevent diffusive transport, we require $R_m\gg1$,
which for our purposes appears to be achieved by $R_m\approx k\gtrsim5$. This
argument also shows that, for the chosen diffusivity profile, the downward
pumping velocity needs to be of the order of 25~m/s. In reality, this pumping
can be due to a mixture of turbulent and topological effects 
and the choice of the form for ${\vec{u_{\mathrm{p}}}}$ is not critical.

\section{Effects of varying the diffusivity and meridional velocity}

In this section we briefly discuss four variations to the above reference case.
Explicitly, we consider one simulation with a potential field boundary
condition, one with a different diffusivity profile, one with anisotropic
diffusivity, and one with a different meridional flow profile.

\subsection{Potential field boundary condition}

\begin{figure}[h]
 \includegraphics[scale=0.5]{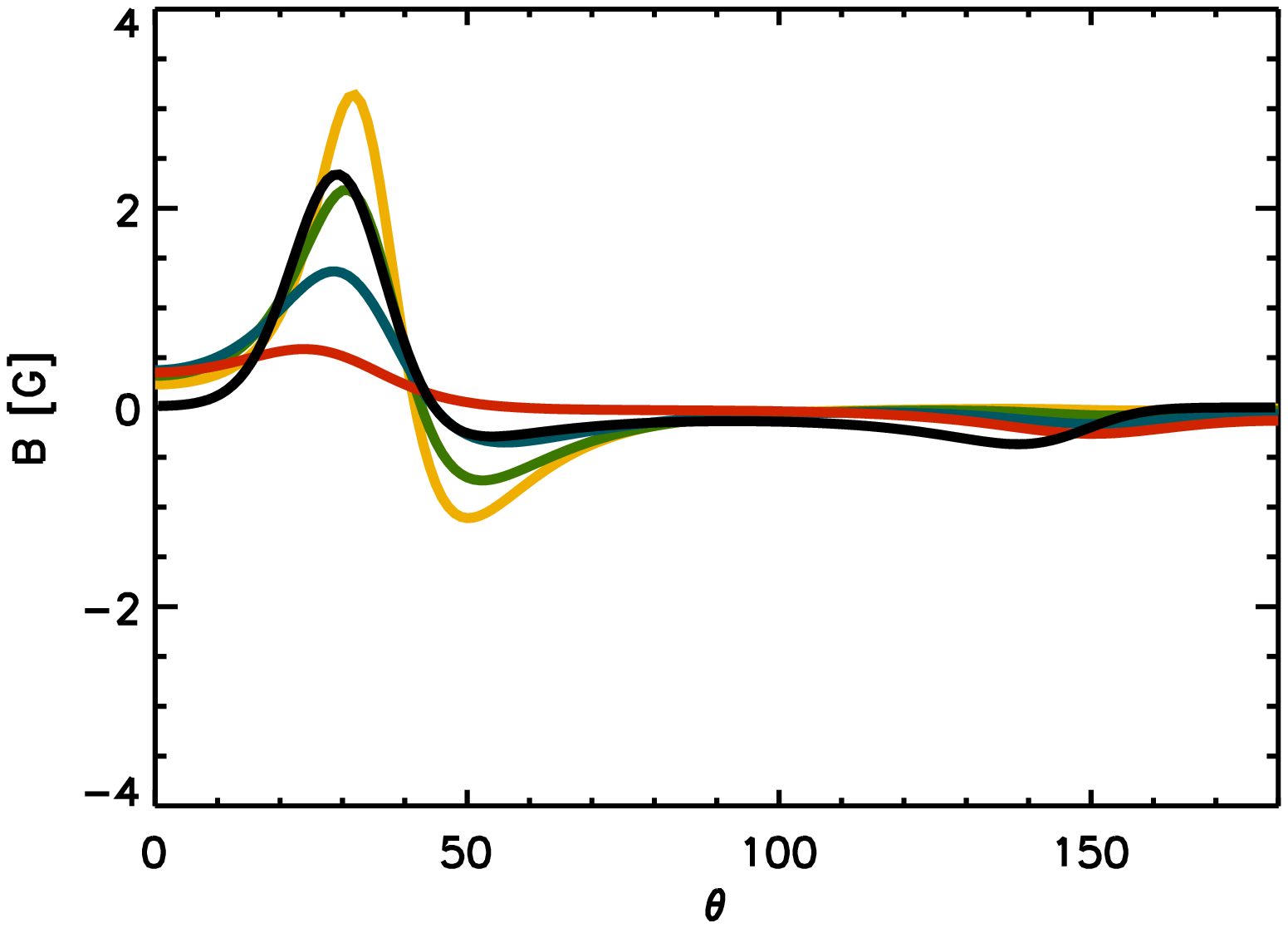}

 \includegraphics[scale=0.5]{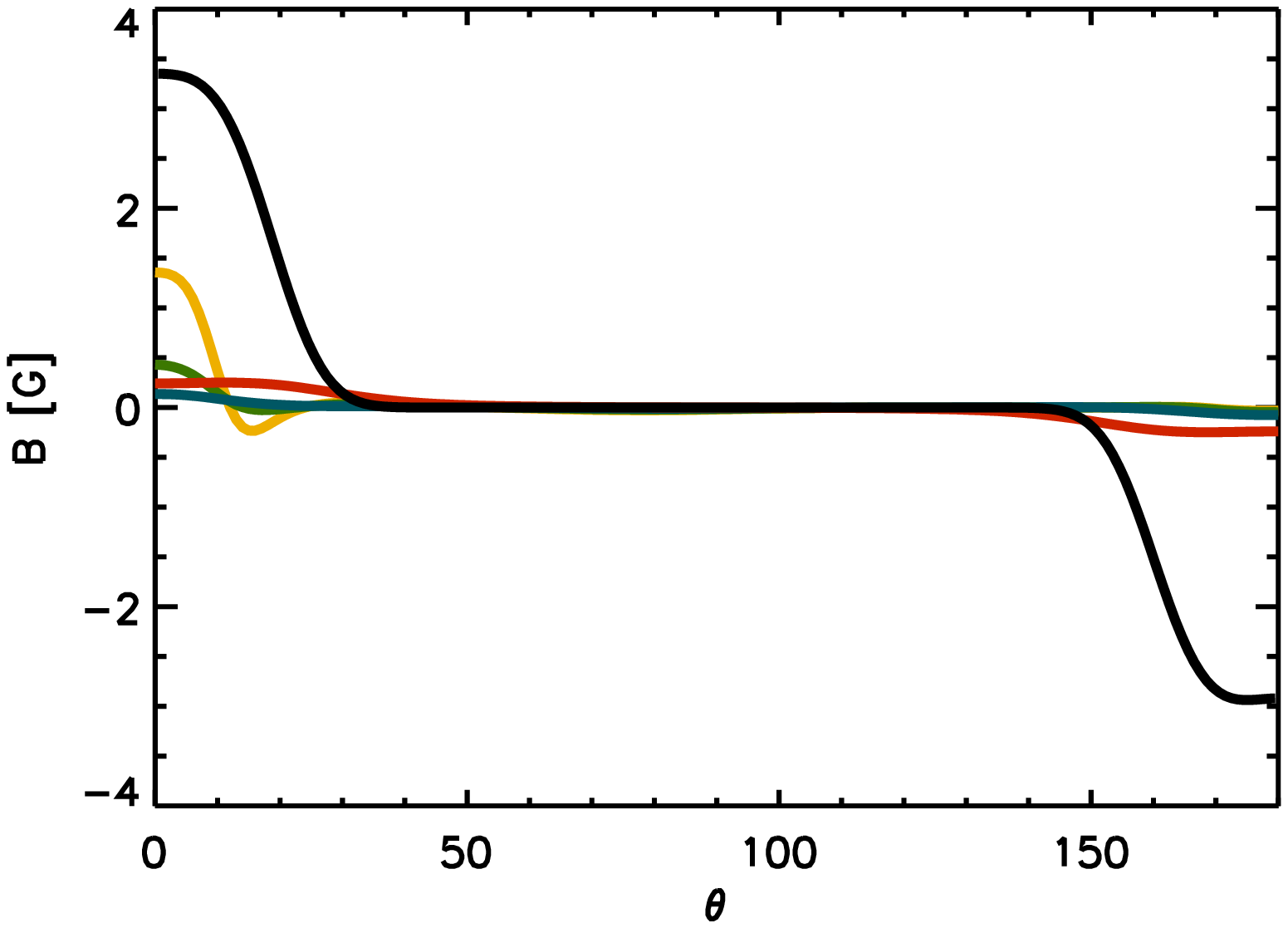}
  \caption{Similar to Fig.~\ref{fig:vert} except a potential field
  upper boundary condition was used for the FTD simulations. The black line shows the surface field
  from the SFT model, the colored lines show the FTD results for different values of $k$.}
  \label{fig:pot}
\end{figure}

\begin{figure}[h]
 \includegraphics[scale=0.9,angle=90]{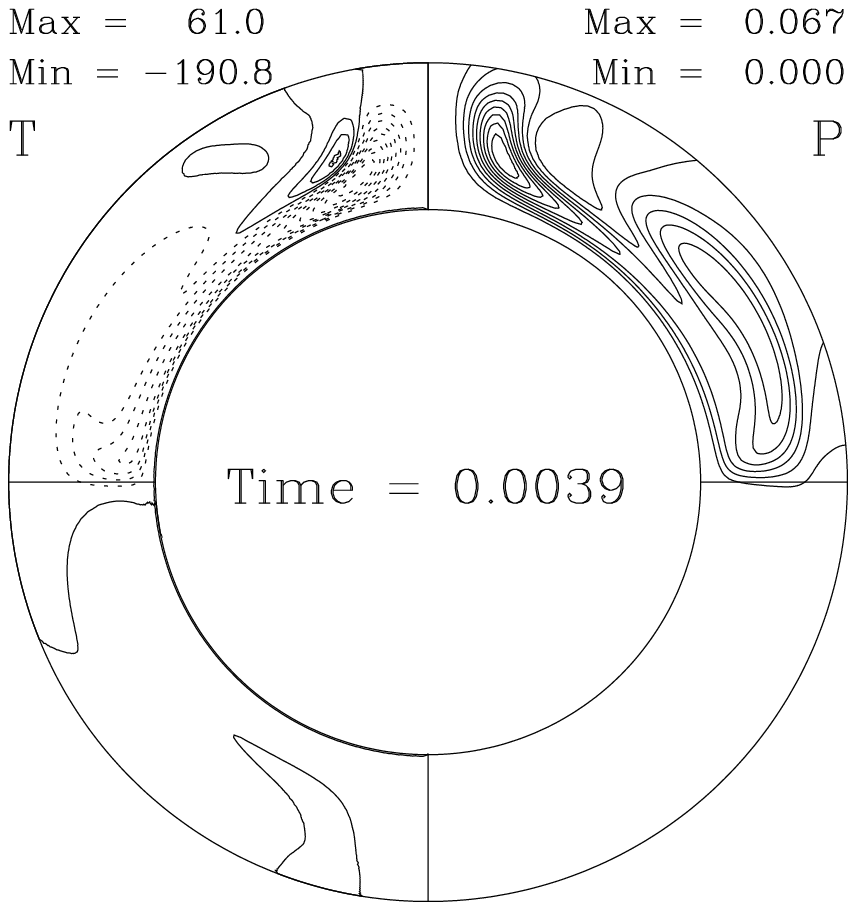}
 
 \includegraphics[scale=0.9,angle=90]{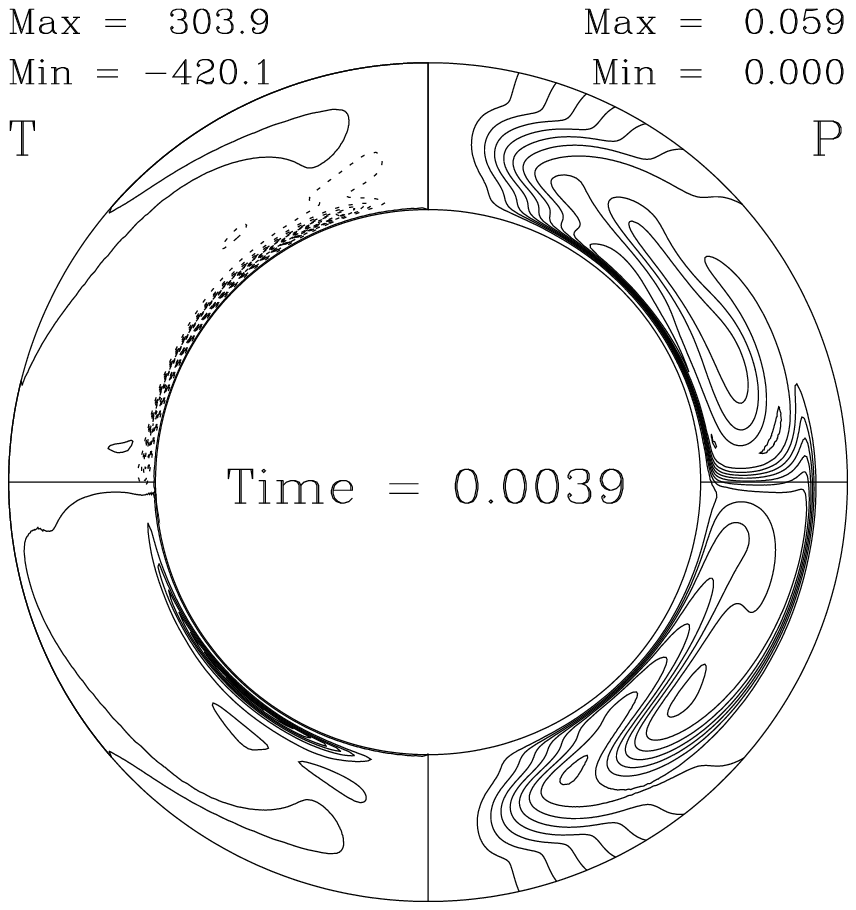}
  \caption{The magnetic field structure from the FTD model in the same format as in Fig.~\ref{fig:fl_vert}
          when the potential field boundary condition is used.}
  \label{fig:fl_pot}
\end{figure}

Figure~\ref{fig:pot} shows the evolution of the field from the FTD simulations
with a potential field boundary condition. The SFT result is again shown for
reference. With this boundary condition we see that the match is always poor.
This is because there is now a strong diffusive flux across the solar surface,
corresponding to the retraction of field lines, as can be seen in
Fig.~\ref{fig:fl_pot}. Hence for the FTD to be consistent with the SFT model,
we need strong downward pumping ($k \gtrsim 5$) and a vertical boundary
condition. These two requirements correspond directly to the assumptions of the
SFT model, that the only sources are those which are explicitly put in (i.e. no
diffusive sources) and that the field at the surface is vertical. We note that
extensions to the SFT model have slightly relaxed the assumption that there are
no diffusive fluxes \citep[see for example][]{Baumann06}, but the values of the
radial diffusivities suggested there correspond to long decay times of the SFT
fields at the poles which are still not comparable to our FTD simulations with
$k=0$, 1 or 2.

\subsection{High diffusivity in the bulk of the convection zone}

For the simulation with a different $\eta$, we considered
$\eta_0=0.1$~km$^2$s$^{-1}$, $\eta_1=100$~km$^2$s$^{-1}$ and
$\eta_2=250$~km$^2$s$^{-1}$ in Eq.~(15). This is similar to the diffusivity
profile of the reference case discussed in Sect.~3 except that the diffusivity
in the bulk of the convection zone has been raised to $100$~km$^2$s$^{-1}$. The
average magnetic diffusivity of the transition between low and high velocities
is then higher, the velocity by contrast has fallen. The magnetic Reynolds
number is then $R_m=(1.5 k)/3.5$. To have $R_m \gtrsim5$ we then need
$k\gtrsim5\times(3.5/1.5)\approx12$; and indeed we found that with $k=10$ the
FTD and SFT models were close to, though not quite, matching.

\subsection{Anisotropic diffusivity}

\begin{figure}[h]
 \includegraphics[scale=0.5]{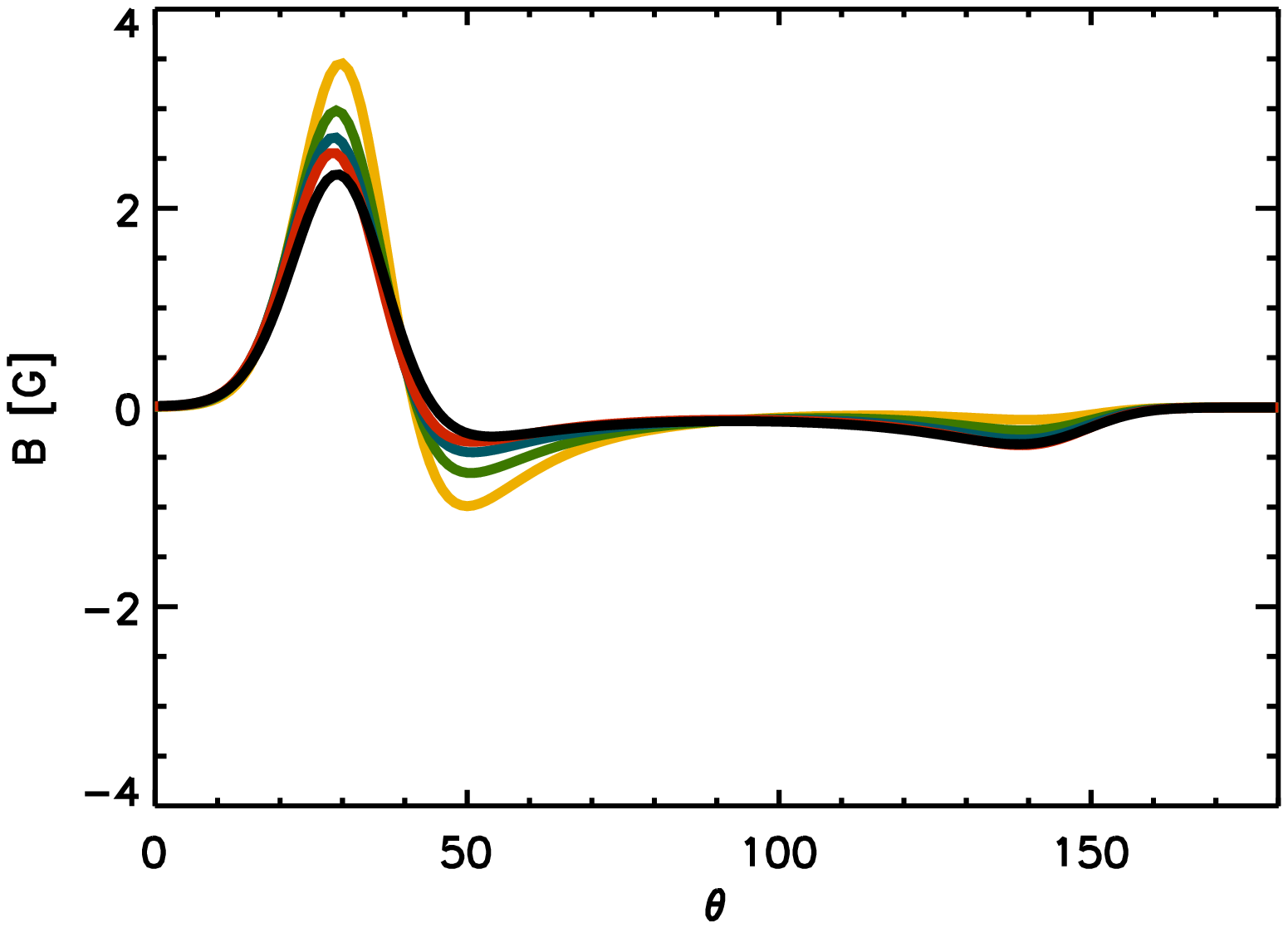}

 \includegraphics[scale=0.5]{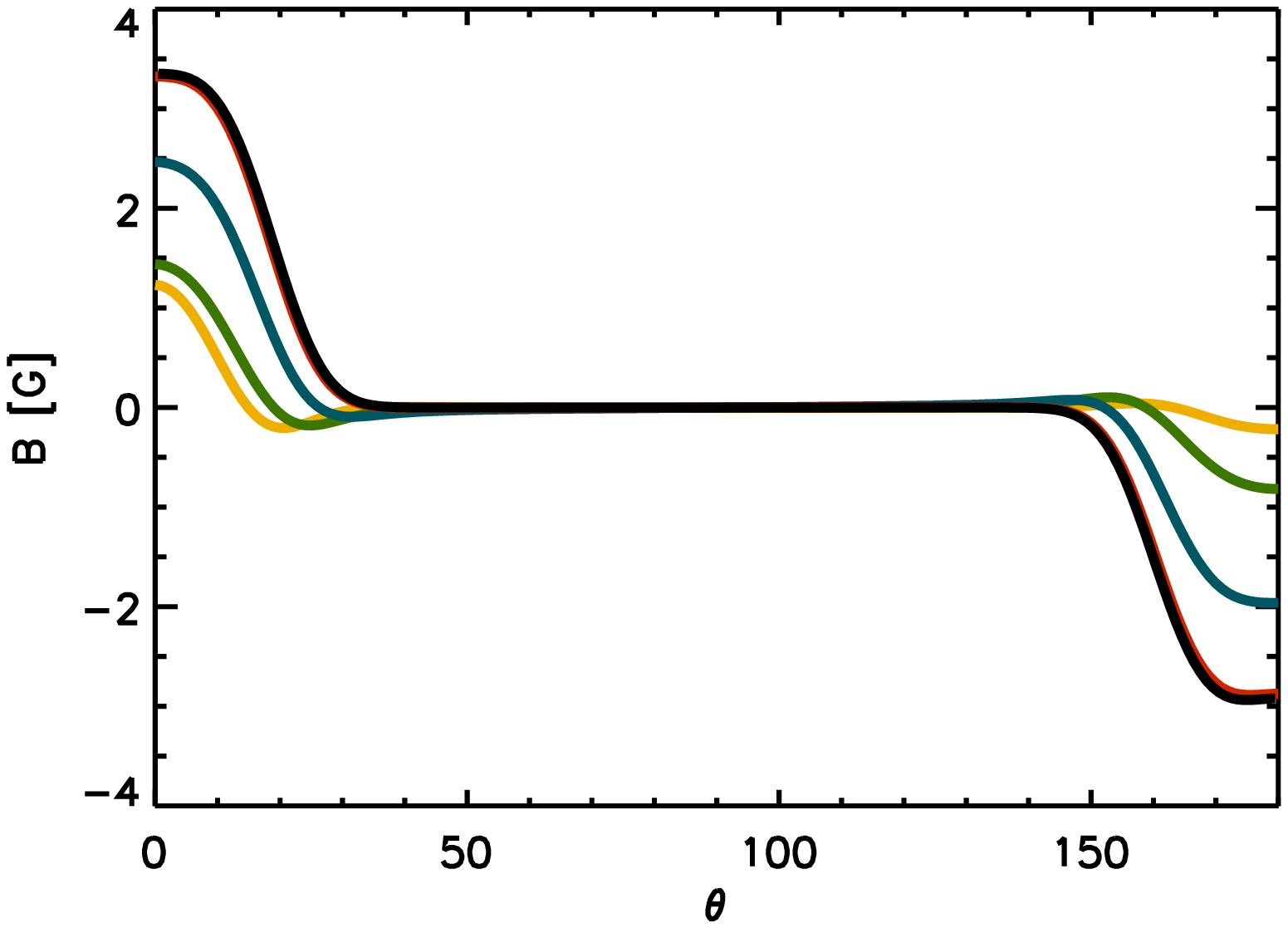}
  \caption{The surface field from the reference SFT simulations at $t=18$ months (top)
   and $t=72$ months (bottom) are shown in black. The results of the FTD simulation with a vertical
   field outer boundary and an anisotropic near-surface diffusivity (250~km$^2$s$^{-1}$ in the
   horizontal directions and 25~km$^2$s$^{-1}$ in the vertical direction). The vertical pumping
   is based on the vertical diffusivity gradient with $k=20$ (red), 10 (blue), 5 (green), and 0 (yellow).}
  \label{fig:anis}
\end{figure}

In our third experiment, we studied the effect of an anisotropy in the
diffusivity near the surface. We used the same formula for the different
components of $\eta$ (i.e., Eq.~\ref{eqn:eta}), but with different values of
the surface diffusivity, $\eta_2$, for the horizontal and vertical directions.
Motivated by the work of \citet{Miesch11}, we chose the longitudinal and
latitudinal diffusivities to be the same, $\eta_2=250$~km$^2$s$^{-1}$, and the
radial diffusivity  to be an order of magnitude smaller,
$\eta_2=25$~km$^2$s$^{-1}$. We based the downward pumping, ${\vec{u_{\mathrm{p}}}}$, on the gradient of
the vertical component of the diffusivity. The comparison of the FTD and SFT
models, for several values of $k$, are shown in Fig.~\ref{fig:anis} for two
times. Importantly, a strong downward pumping with $k>10$ is needed for the FTD
to match the SFT surface evolution.

\subsection{Variation of the meridional circulation}

\begin{figure}[h]
 \includegraphics[scale=0.5]{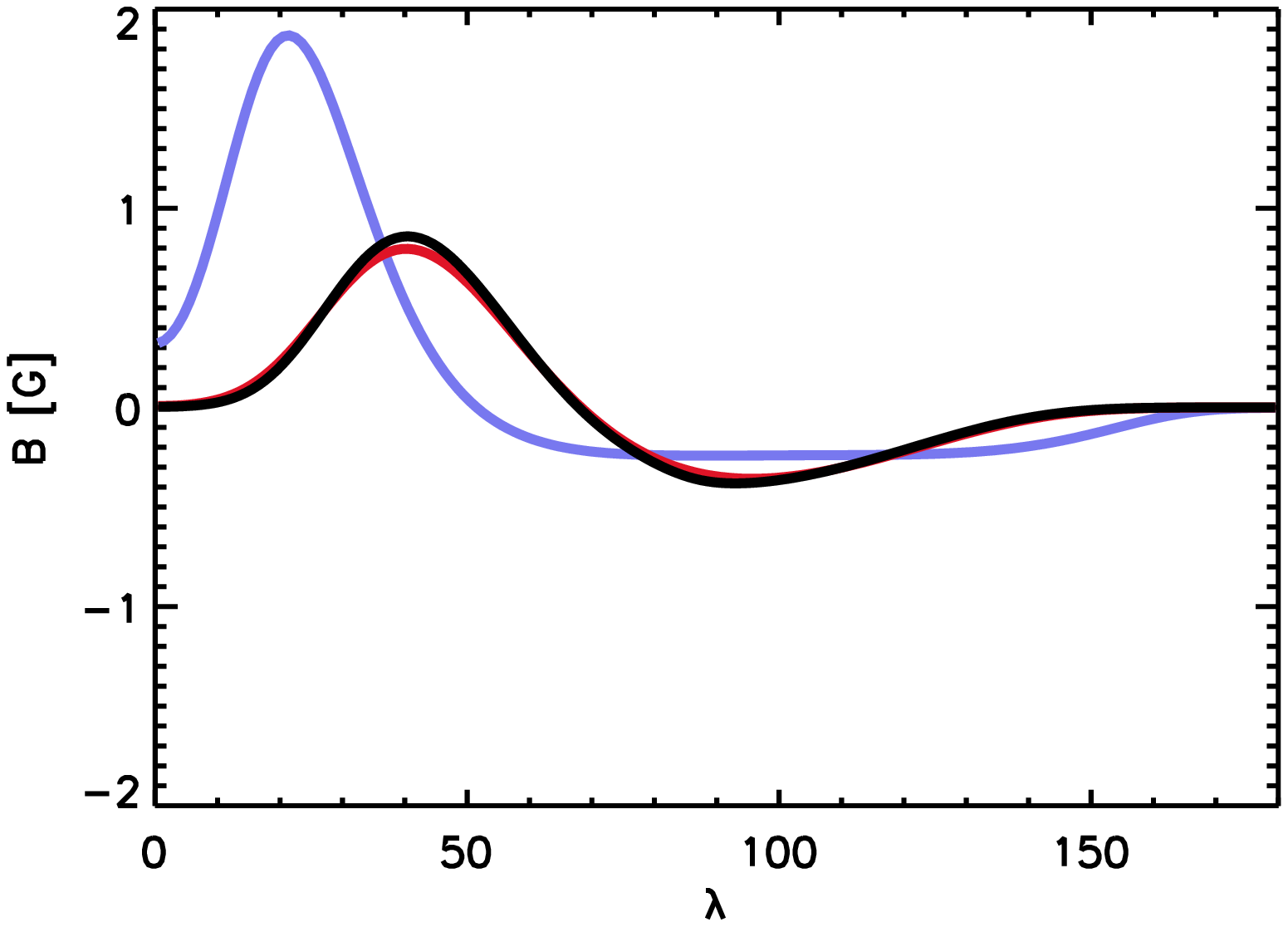}

 \includegraphics[scale=0.5]{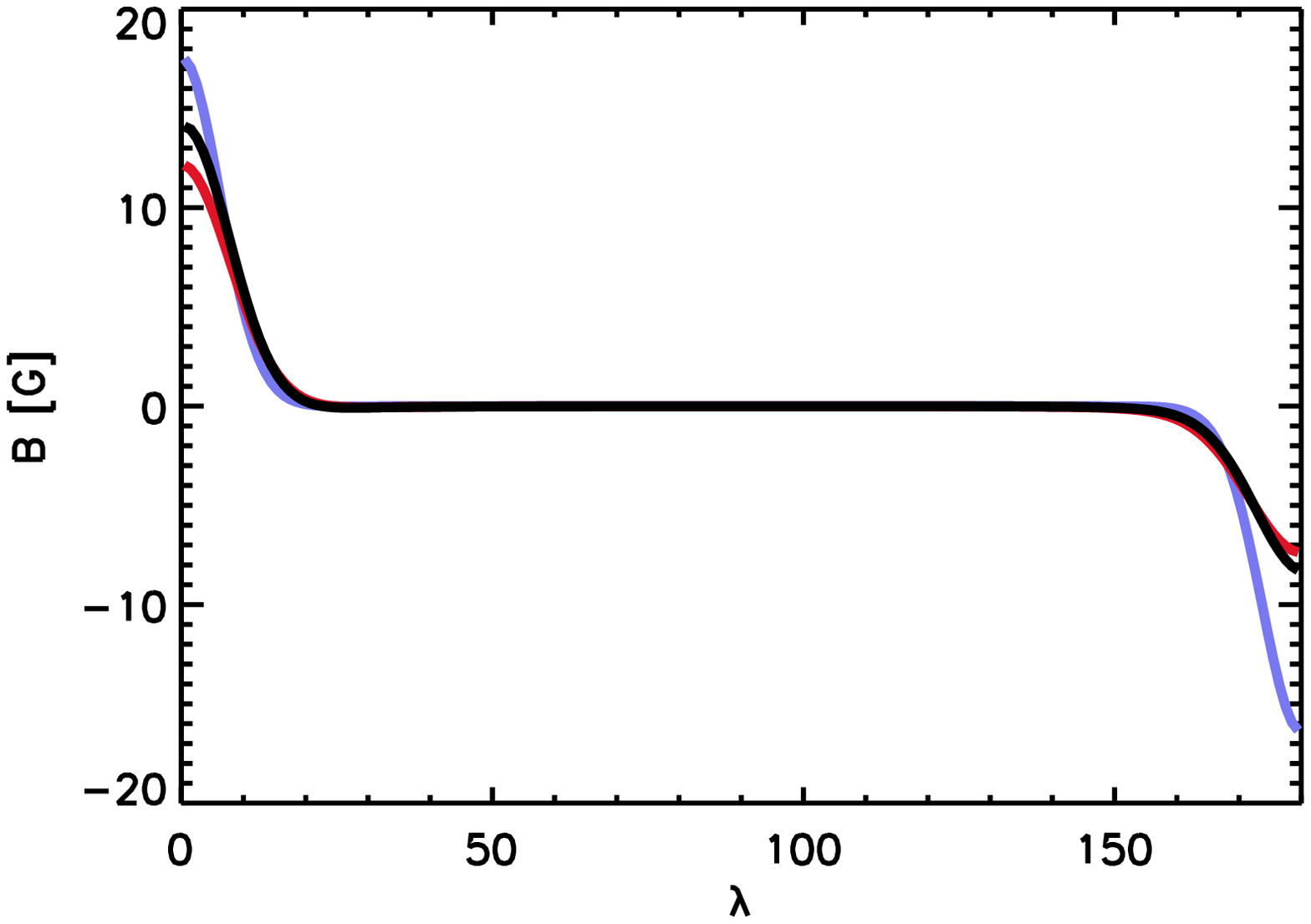}
  \caption{The surface field from FTD simulations at $t=18$ months (top)
   and $t=72$ months (bottom) with a vertical
   field outer boundary using the $k=5$ and the meridional velocity profile
  with $p=0.25$, $q=0$ and $m=0.5$ (black) is used. In this case there is a strong near-surface shear.
  The results from the SFT model  using the surface meridional velocity (blue) and using the average of
  the meridional velocity above 0.95 $R_{\sun}$ (red) are shown for comparison.}
  \label{fig:mv2}
\end{figure}

For the simulation with a different meridional velocity profile, we used the
same form as described in Eqs.~(\ref{eqn:ur_start}) to (\ref{eqn:ur_end}) but
with $p=0.25$, $q=0$ and $m=0.5$ \citep{Dikpati99}. For this choice of the
meridional flow, the FTD and SFT models always evolve differently, even though
the surface velocity is used for the SFT calculation (Fig.~\ref{fig:mv2},
top). The reason is that the meridional velocity in this case is not constant
above the transition from low to high diffusivities, which occurs at about
0.95~$R_{\sun}$. The magnetic field in the FTD calculation sees a range of
velocities above the `boundary layer' associated with the transition and the
strong pumping. Because the diffusivity is reasonably large above the
transition, the magnetic flux should essentially be advected according to the
average meridional flow in this layer. Therefore the surface field is
effectively advected with the average meridional flow speed in the boundary
layer, and not with its surface value. This indeed happens as can be seen in
Fig.~\ref{fig:mv2} (bottom). It is noteworthy that this mainly affects the time
it takes for the flux to reach the poles, not the amount that eventually gets
there. The meridional flow is difficult to measure at depths below about 10~Mm;
in the top 10~Mm the indications from helioseismology are that the meridional
flow first increases and then decreases \citep{Basu10}.

There is also an observed near-surface shear in the differential rotation
\citep{Thomson95}, which has been used to explain the observed difference
between the rotation of magnetic features \citep{Snodgrass83} and the rate
deduced from surface Doppler observations of the flow. For a review of the
observational results, see \citet{Beck00}. The conventional explanation is made
in terms of the `anchoring depth' of the features \citep{Nesme-Ribes93}. Our
suggestion is that the observed rotation rate of magnetic features is 
partly due to the average value in a high-diffusivity layer, which is partially isolated
from the deeper dynamics by a boundary layer associated with magnetic pumping.

\section{Conclusion}

With a vertical outer boundary condition and enough pumping the FTD model is
consistent with the SFT model. The pumping needs to be strong enough to result
in a magnetic Reynolds number of approximately 5. With a potential boundary
condition or weaker pumping, the models do not match. This strong pumping
requires a velocity which is greater than the standard value given by
mean-field theory for  diamagnetic pumping. Since the SFT model matches
observations, it follows that the vertical field boundary condition and
sufficient downward pumping are required for the FTD model to match the
observed surface evolution of the field. 

\begin{acknowledgements}
The authors gratefully acknowledge Manfred Sch\"ussler for enlightening
discussions on various aspects of this paper. JJ acknowledges financial support
from the National Natural Science Foundations of China through grants 11173033,
11178005, 11125314 and a Young Researcher Grant of the National Astronomical
Observatories, Chinese Academy of Sciences.
\end{acknowledgements}

\bibliographystyle{aa}
\bibliography{FTD}

\end{document}